\documentstyle[preprint,aps]{revtex}
\begin{document}
\draft
\preprint{Phys. Rev. Lett. {\bf 74}, 4400 (1995)}
\title{
A new cross term in the two-particle\\
Hanbury-Brown-Twiss correlation function
}
\author{Scott Chapman, Pierre Scotto and Ulrich Heinz}
\address{
Institut f\"ur Theoretische Physik, Universit\"at Regensburg,\\
D-93040 Regensburg, Germany
}
\maketitle
\begin{abstract}
Using two specific models and a model-independent formalism, we show
that in addition to the usual quadratic ``side'', ``out'' and
``longitudinal'' terms, a previously neglected ``out-longitudinal''
cross term arises naturally in the exponent of the two-particle
correlator. Since its effects can be easily observed, such a term
should be included in any experimental fits to correlation data.  We
also suggest a method of organizing correlation data using rapidity
rather than longitudinal momentum differences since in the former
every relevant quantity is longitudinally boost invariant.
\end{abstract}

\pacs{25.70.Pq}

\narrowtext

The experimentally measured Hanbury-Brown Twiss (HBT) correlation
between two identical particles emitted in a high energy collision
defines a six dimensional function of the momenta ${\bf p}_1$ and
${\bf p}_2$ \cite{hbt}. A popular way of presenting these is in terms
``size parameters" derived from a gaussian fit to the data of the form
\cite{na35,fixed,both,lcms}
 \begin{equation}
   C({\bf q},{\bf K}) = 1 \pm \lambda \exp \left[
                      - q_s^2 R_s^2({\bf K})
                      - q_o^2 R_o^2({\bf K})
                      - q_l^2 R_l^2({\bf K}) \right]
 \label{0}
 \end{equation}
where ${\bf q}={\bf p}_1-{\bf p}_2$, ${\bf K} = {\textstyle
\frac{1}{2}}({\bf p}_1 +{\bf p}_2)$, the $+$ ($-$) sign is for bosons
(fermions), and the HBT cartesian coordinate system is defined as
follows \cite{bertsch2}: The ``longitudinal'' or $\hat{z}$ (subscript
$l$) direction is parallel to the beam; the ``out'' or $\hat{x}$
(subscript $o$) direction is parallel to the component of ${\bf K}$
which is perpendicular to the beam; and the ``side'' or $\hat{y}$
(subscript $s$) direction is the remaining direction.

In this letter we assert that significantly more can be learned and
better fits achieved if an ``out-longitudinal'' cross term is included
in any gaussian fits to the data.  In other words, we suggest that the
data should be fit to a function with the following form
 \begin{equation}
  C({\bf q},{\bf K}) =  1\pm \lambda\exp\left[
  - q_s^2 R_s^2({\bf K}) - q_o^2 R_o^2({\bf K})
  - q_l^2 R_l^2({\bf K}) - 2 q_o q_l R_{ol}^2({\bf K}) \right]\;,
 \label{6g}
 \end{equation}
where $R_{ol}^2$ is a parameter which can be either positive or
negative; we simply use the $R^2$ notation to denote the fact that it
is has the dimension of an area.

Since particles 1 and 2 are indistinguishable, the overall sign of
${\bf q}$ is irrelevant. The relative signs of the various components
of ${\bf q}$, however, are well defined physical quantities for any
given pair. Our sign convention for ${\bf q}$ will be such that $q_s$
is always positive. We can thus unambiguously discuss correlations for
negative as well as positive values of both $q_o$ and $q_l$.

To see how an ``out-longitudinal'' cross term arises in two-particle
correlations, we use the following well established theoretical
approximation \cite{pratt,chap2}
 \begin{equation}
  C({\bf q}, {\bf K}) \simeq 1 \pm \frac{|\int d^4x\,
  S(x,K)\,e^{iq{\cdot}x}|^2} {|\int d^4x\, S(x,K)|^2}
 \label{6}
 \end{equation}
where $q_0 = E_1-E_2$, $K_0=E_K=\sqrt{m^2+|{\bf K}|^2}$. Here
$S(x,K)$ is a function which describes the phase space density of the
emitting source. For pairs with $|{\bf q}| \ll E_K$
 \begin{equation}
   q{\cdot}x \simeq (\beta_o q_o + \beta_l q_l)\,t - q_o x
  -q_s y -q_l z \;,
 \label{6a}
 \end{equation}
where $\beta_i=K_i/E_K$.

As a simple example, we consider the following cylindrically symmetric
gaussian emission function
 \begin{equation}
  S(x,K) = f(K)\, \exp \left[-\frac{x^2+y^2}{2 R^2}-\frac{z^2}{2 L^2}
  -\frac{(t-t_0)^2}{2(\delta t)^2} \right]\;.
 \label{6b}
 \end{equation}
Using (\ref{6}) and (\ref{6a}), it is easy to see that the corresponding
correlation function takes the form:
 \begin{equation}
  C({\bf q},{\bf K})
  = 1 \pm \exp \left[-q_s^2 R^2 - q_o^2 (R^2+\beta_o^2(\delta t)^2)
  - q_l^2 (L^2 + \beta_l^2(\delta t)^2)
   - 2 q_o q_l \beta_o \beta_l (\delta t)^2 \right]\;.
 \label{6c}
\end{equation}
For this model, the $q_oq_l$ cross term thus provides a measurement of
the duration of particle emission $(\delta t)$.

One might think that this cross term is just a trivial kinematic effect
which would not arise if the correlation were calculated in some more
carefully chosen coordinate system or reference frame.  For example,
for spherically symmetric systems ($L=R$), it has been shown that the
cross term vanishes if axes are chosen parallel and perpendicular to
${\bf K}$ (rather than parallel and perpendicular to the beam)
\cite{pratt2}.  The reader can verify, however, that for systems with
$L\ne R$, the cross term does not vanish in these rotated coordinates,
but rather measures the $L^2-R^2$ asymmetry of the source.

Another system that is often proposed is the Longitudinally Co-Moving
System (LCMS), which is defined as the frame in which $\beta_l=0$
\cite{csorgo3,both,lcms}.  Glancing at (\ref{6c}), it naively appears
that the cross term will vanish in this frame. Being more careful,
however, one can see that after transforming to the LCMS frame (primed
variables)
\begin{equation}
  t^\prime = \gamma_l(t-\beta_l z)\;\;\;\;\;\;\;\;\;
  z^\prime =\gamma_l(z-\beta_l t)\;,
 \label{6d}
 \end{equation}
where $\gamma_l = 1/\sqrt{1-\beta_l^2}$,\  $t^\prime z^\prime$
terms arise in the transformed emission function $S^\prime$. These in
turn lead to a nonvanishing cross term of the form
 \begin{equation}
  R_{ol}^{\prime 2} =
  \beta_o\beta_l\gamma_l^2 \left[(\delta t)^2 + L^2\right]\;,
 \label{6d1}
\end{equation}
where $\beta_l$ and $\gamma_l$ are evaluated in the CM frame. We can
see that the cross term cannot in general be removed simply by
transforming to another coordinate system or reference frame, and
therefore it is certainly not just a trivial kinematic effect. In
fact, $R_{ol}^2$ contains physical information about the emitting source
which is just as important as that found from evaluating the
difference $R_o^2-R_s^2$.

In order to get a broader feeling for what the cross term measures, we
now introduce a general formalism valid for any cylindrically
symmetric emission function which can be expressed in a roughly
gaussian form. Using (\ref{6a}), we expand $\exp(iq{\cdot}x)$ in
(\ref{6}) for $q{\cdot}x\ll 1$ to find (see \cite{bertsch})
 \begin{eqnarray}
  C({\bf q}, {\bf K}) = 1 \pm \biggl\{1 &-& q_s^2\left\langle
  y^2\right\rangle
  -\left\langle\left[q_o\left(\beta_o t - x\right) +
  q_l\left(\beta_l t-z\right)\right]^2\right\rangle
\nonumber \\
  &+& \Bigr\langle q_o\left(\beta_o t - x\right) +
  q_l\left(\beta_l t-z\right)\Bigr\rangle^2
  + {\cal O}\left[(q{\cdot}x)^4\right]\biggr\}\;,
 \label{6e}
 \end{eqnarray}
where the $q_s q_o$ and $q_s q_l$ terms vanish due to cylindrical
symmetry \cite{chap2} and we have introduced the
notation
 \begin{equation}
  \langle \xi \rangle \equiv \langle \xi \rangle ({\bf K}) =
  \frac{\int d^4x\, \xi\, S(x,K)}{\int d^4x\, S(x,K)}\;.
 \label{6f}
 \end{equation}

Exponentiating (\ref{6e}), we can see that for any cylindrically
symmetric system, the correlation function for small momentum
differences $(q_iR_i\ll 1)$ can be expressed in the form of Eq.
(\ref{6g}) where $\lambda=1$ and the functions $R_i^2({\bf K})$ can
simply be read off as the coefficients of the corresponding $q_iq_j$
terms in Eq.~(\ref{6e}):
 \begin{eqnarray}
   R_s^2
   & = & \left\langle y^2\right\rangle - \left\langle y\right\rangle^2
     =   \left\langle y^2\right\rangle
 \nonumber \\
   R_o^2
   & = & \left\langle(x-\beta_o t)^2\right\rangle
        -\langle x-\beta_o t\rangle^2
 \nonumber \\
   R_l^2
   & = & \left\langle(z-\beta_l t)^2\right\rangle
       - \langle z-\beta_l t\rangle^2
 \nonumber \\
   R_{ol}^2
   & = & \left\langle(x-\beta_o t)(z-\beta_l t)\right\rangle
       - \langle x-\beta_o t\rangle \langle z-\beta_l t\rangle \, .
 \label{6h}
 \end{eqnarray}
These expressions have the typical form of variances and show that the
HBT ``size parameters" are really lenghts of homogeneity associated
with the source function $S(x,K)$ \cite{sinyu3}. For a static source,
these homogeneity lengths are equal to its geometric size in the
various directions which can then be directly extracted from the HBT
correlator. For expanding sources the interpretation of the HBT size
parameters is more involved
\cite{pratt,chap2,pratt2,csorgo3,sinyu3,csorgo}, and the HBT
parameters are usually smaller than the geometric extensions of the
source. The cross term is seen to measure the temporal extent of the
source as well as the $x\,z$, $x\,t$, and $z\,t$ correlations of the
emission function. Note that the LCMS radii can be found from
(\ref{6h}) by setting $\beta_l=0$ and using $S(x',K')$ (see
Eq.~(\ref{6d})). The cross term vanishes in this frame if and only if
the source $S(x',K')$ is reflection symmetric under $z' \to - z'$
(which is not the case for our source (\ref{6b})).

One might argue that the model independent expressions of Eq.
(\ref{6h}) should not be compared to experimental correlation radii
since the former measure second derivatives of the correlation
function around ${\bf q}=0$ (because we used $q{\cdot}x\ll 1$ to
derive them), while the latter are parameters of a gaussian fit to the
whole correlation function \cite{na35,fixed,both,lcms}.  On the other
hand, for any source which has a roughly gaussian profile in some
complete set of spatial coordinates, the two different methods of
measuring radii will give roughly the same results.  For these types
of models, the simple expressions generated by Eq. (\ref{6h}) provide
valuable insights as to how various parameters of the source
distribution qualitatively affect measurable features of the
correlation function.

We already discussed one gaussian model in Eq.~(\ref{6b}), but here
we would like to discuss another, possibly more realistic model which
is similar to the ones presented in \cite{csorgo}. In the center of
mass frame of an expanding fireball, we define the following emission
function
 \begin{equation}
  S(x,K)
  = \frac{\tau_0 m_t {\rm ch}(\eta-Y)}
           {(2\pi)^3 \tau \sqrt{2\pi(\delta \tau)^2}}
      \exp \left[- \frac{K{\cdot}u(x)}{T}
  - \frac{\rho^2}{2R_G^2}
                        - \frac{\eta^2}{2(\delta \eta)^2}
                        - \frac{(\tau-\tau_0)^2}{2(\delta \tau)^2}
                  \right]\, ,
 \label{2.2}
 \end{equation}
where $T$ is a constant freeze-out temperature, $\rho = \sqrt{x^2 +
y^2}$, $\tau=\sqrt{t^2-z^2}$,
\newline
$\eta = {\textstyle \frac{1}{2}}
\ln[(t+z)/(t-z)]$, $m_t = \sqrt{m^2 + K_\perp^2}$, and $Y$ is the
rapidity of a particle with momentum ${\bf K}$. Note that in the limit
$\delta \tau \to 0$, (\ref{2.2}) becomes the Boltzmann approximation
for a hydrodynamic system with local flow velocity $u(x)$ which
freezes out on a 3-dimensional hypersurface of constant longitudinal
proper time $\tau_0$ and temperature $T$ \cite{marb,chap2}. We will
consider a flow which is non-relativistic transversally but which
exhibits Bjorken expansion longitudinally
 \begin{equation}
   u(x) \simeq
   \biggl(\left(1+{\textstyle\frac{1}{2}}(v\rho/R_G)^2
           \right){\rm ch}\eta, \,(vx/R_G),
   (vy/R_G),\,
      \left(1+{\textstyle\frac{1}{2}}(v\rho/R_G)^2\right)
      {\rm sh}\eta\,\biggr)\;,
 \label{2.1}
 \end{equation}
where $v\ll 1$ is the transverse flow velocity of the fluid at $\rho =
R_G$. In \cite{chap2} we show that when the emission function
(\ref{2.2}) is integrated over spacetime, it produces a very
reasonable one-particle distribution.

To facilitate calculating the correlation function, we can make the
physically reasonable assumption that $\delta\tau/\tau_0\,
{\mathrel{\lower.9ex\hbox{$\stackrel{\displaystyle <}{\sim}$}}}
\,{\textstyle\frac{1}{2}}$ so that we can be justified in replacing
integrals over only positive values of $\tau$ with ones ranging from
$-\infty$ to $+\infty$. We can then achieve analytic results by
making a Taylor expansion \cite{chap2} in the parameter
 \begin{equation}
   \frac{1}{(\delta\eta)_*^2}
   = \frac{1}{(\delta\eta)^2} + \frac{m_t}{T}\;.
 \label{2.5a}
 \end{equation}
Note that for pairs in which $m_t/T \gg 1/(\delta\eta)^2$ as were
studied in \cite{sinyu2}, $(\delta\eta)_*^2$ becomes simply $T/m_t$.

Using the expressions (\ref{6h}) and keeping a few subleading
corrections which are important when considering pions \cite{chap2},
we find
 \begin{eqnarray}
   R_s^2 & = & R_*^2\;;
 \nonumber \\
   R_o^2 & = & R_*^2 + \frac{K_\perp^2}{m_t^2}
   (\delta \tau)^2 + \frac{K_\perp^2}{m_t^2}\beta_l^2
   \tau_0^2 (\delta\eta)_*^2
      + \frac{K_\perp^2}{m_t^2} \left[1+\beta_l^2
      - 2 \beta_l\frac{Y}{(\delta\eta)^2}\right]
        (\delta \tau)^2(\delta\eta)_*^2
\nonumber \\
	& & \phantom{R_*^2}
      + \frac{K_\perp^2}{m_t^2}\tau_0^2\left[\beta_l^2 \nu
      - 2 \beta_l \frac{Y}{(\delta\eta)^2}
      + {\textstyle\frac{1}{2}}\right](\delta\eta)_*^4\;;
 \nonumber \\
   R_l^2 & = & \frac{m_t^2}{E_K^2}
   \tau_0^2 (\delta\eta)_*^2 + \frac{m_t^2}{E_K^2}(\delta\tau)^2
   (\delta\eta)_*^2+ \frac{m_t^2}{E_K^2} \nu \tau_0^2 (\delta\eta)_*^4\;;
 \nonumber \\
   R_{ol}^2 & = &
   -\beta_o\beta_l\tau_0^2 (\delta\eta)_*^2 -\beta_o \left[
   \beta_l -\frac{Y}{(\delta\eta)^2}\right](\delta\tau)^2(\delta\eta)_*^2
    - \beta_o\tau_0^2 \left[\beta_l \nu
      - \frac{Y}{(\delta\eta)^2}\right](\delta\eta)_*^4\;.
 \label{2.6}
 \end{eqnarray}
where
 \begin{equation}
   \frac{1}{R_*^2} = \frac{1}{R_G^2}\left(1 + \frac{m_t}{T}v^2\right)
 \label{2.5}
 \end{equation}
and $\nu = 1 + (R_*/R_G)^2 -{\textstyle \frac{1}{2}} (m_t/T)
(\delta\eta)_*^2$. As pointed out in \cite{csorgo} and seen from Eq.
(\ref{2.5}), transverse flow causes the ``side'' radius to measure
something smaller than the real geometrical radius $R_G$.  What it
does measure is the transverse region of homogeneity of the fluid as
seen by particles with a given $p_t$ \cite{sinyu3,chap2}. It is also
interesting that $R_o^2-R_s^2$ depends on the average rapidity and is
not quite directly proportional to the duration of particle emission
$\delta\tau$ even for pairs with $\beta_l=0$ \cite{time}. In our
opinion, however, the most interesting feature of this model is the
cross term radius $R_{ol}^2$. Although just as in Eq.~(\ref{6c}) the
cross term vanishes for pairs with either $\beta_o=0$ or $\beta_l=0$,
it will in general have an important effect on the correlation
function, especially for pairs with large $|Y|$.

The effect becomes most easily apparent when we plug in some numbers
and plot the correlation function. For simplicity, we consider a pion
source with no transverse flow ($v=0$) which freezes out
instantaneously ($\delta\tau=0$) with the following other source
parameters: $R_G=3$ fm, $\tau_0=4$ fm/c, $\delta\eta = 1.5$, and
$T=150$ MeV. Restricting ourselves to pairs with $Y=-2$, $K_\perp=200$
MeV, and $q_s=0$, we can now calculate the correlation function both
by using the approximate analytic radii of (\ref{2.6}) and by
performing a numerical calculation directly from Eqs. (\ref{6}) and
(\ref{2.2}). Comparing the results, we have found that the gaussian
approximation of (\ref{6g}) with the radii of (\ref{2.6}) is able to
describe the numerically calculated correlation function to within
about 20\% \cite{chap2}. Figure 1 shows a plot of the latter as a
function of $q_o$ and $q_l$. The effect of the cross term can be seen
in the form of an asymmetric ridge running from the peak at
$q_o=q_l=0$ down to the front left where $q_l>0$ and $q_o<0$.  This
kind of ridge is clearly identifiable experimentally and has in fact
already been seen in preliminary NA35 correlation data \cite{alber}.

Due to the boost invariance of the flow profile, the LCMS radii
corresponding to the above model can be obtained simply by setting
$\beta_l=0$ and $E_K=m_t$ in (\ref{2.6}). Note that the factor $Y$ in
the second line of $R_{ol}^2$ should not be set equal to zero, since
it arises from the $\eta$ distribution of the source (\ref{2.2}) which
obviously breaks the boost invariance of the emission function. It can
be verified that if a source is {\em completely} boost invariant then
the cross term will always vanish in the LCMS frame \cite{sinyu2}.
However, any source with a finite size ($\delta\eta < \infty$) cannot
be completely boost invariant, so it will in general feature a
nonvanishing cross term in the LCMS, since the LCMS does not
coincide with the local rest frame \cite{chap3}.

We would now like to suggest a better way of organizing correlation
data from sources undergoing boost-invariant longitudinal expansion
\cite{aver}. Returning to (\ref{6}), let us make an alternative
on-shell definition of the 4-vector $K$:
 \begin{equation}
   K=(m_t{\rm ch}Y,{\bf K}_t,m_t{\rm sh}Y)
 \label{2.8}
 \end{equation}
where ${\bf K}_t={\textstyle\frac{1}{2}}({\bf p}_{1t}+{\bf p}_{2t})$,
$m_t^2=m^2+|{\bf K}_t|^2$, $Y={\textstyle\frac{1}{2}}({\rm y}_1+{\rm
y}_2)$, and ${\rm y}_i$ is the rapidity of the $i$th particle. The
resulting approximation is at least as good as the approximation we
have been using up to now \cite{chap2}. This definition suggests that
we express the correlation function in terms of $q_s$, $q_o$ and the
rapidity difference ${\rm y}={\rm y}_1-{\rm y}_2$:
 \begin{equation}
   C({\rm y},q_s,q_o,Y,K_\perp) \simeq
    1 \pm\lambda \exp[-q_s^2R_s^2 - q_o^2R_o^2 -{\rm y}^2\alpha^2 -
      2 q_o {\rm y} R_{o {\rm y}}]\, .
 \label{2.9}
 \end{equation}
The reader should take care not to confuse the rapidity difference
${\rm y}$ with the cartesian coordinate $y$.

The model independent expressions corresponding to (\ref{6h}) are now
given by:
 \begin{eqnarray}
   R_s^2 & = & \langle y^2\rangle
 \nonumber \\
   R_o^2 & = & \left\langle\left[x-(K_\perp/m_t)\tau{\rm  ch}(\eta-Y)
               \right]^2\right\rangle
    - \Bigl\langle x-(K_\perp/m_t)\tau{\rm ch}(\eta-Y)
        \Bigr\rangle^2
 \nonumber \\
   \alpha^2 & = & \left \langle\left[m_t\tau{\rm sh}(\eta-Y)\right]^2
                  \right\rangle
                 -\left \langle m_t\tau{\rm sh}(\eta-Y)\right\rangle^2
 \nonumber \\
   R_{o {\rm y}} & = & \Bigl\langle\left[m_t x
          - K_\perp\tau{\rm ch}(\eta-Y) \right]\tau{\rm sh}(\eta-Y)
          \Bigr\rangle
     - \Bigl\langle m_t x-K_\perp\tau{\rm ch}(\eta-Y)
         \Bigr\rangle
         \Bigr\langle\tau{\rm sh}(\eta-Y)\Bigr\rangle\, .
 \label{2.9a}
 \end{eqnarray}
For the model (\ref{2.2}), the radii take the much simpler
form:
 \begin{eqnarray}
   R_s^2 & = & R_*^2
 \nonumber \\
   R_o^2 & = & R_*^2 + \frac{K_\perp^2}{m_t^2}\Bigl[
   \left(1+(\delta\eta)_*^2\right)(\delta \tau)^2
   +{\textstyle\frac{1}{2}}(\delta\eta)_*^4\tau_0^2\Bigr]
 \nonumber \\
   \alpha^2 & = & m_t^2(\delta\eta)_*^2
   \Bigl[\tau_0^2\left(1 + \nu(\delta\eta)_*^2\right)
   +(\delta\tau)^2\Bigr]
 \nonumber \\
   R_{o {\rm y}} & = &
   \frac{K_\perp Y}{(\delta\eta)^2}(\delta\eta)_*^2
   \left[(\delta\eta)_*^2\tau_0^2 +(\delta\tau)^2\right]\;.
 \label{2.9b}
 \end{eqnarray}
The astute reader will note that the above ``side'' and ``out'' radii
are identical to the LCMS versions of (\ref{2.6}), and that aside from
a slight difference in the definition of $Y$, $R_l({\rm LCMS}) =
\alpha/m_t$ and $R_{ol}^2({\rm LCMS})=R_{o {\rm y}}/m_t$. In fact,
for systems undergoing Bjorken longitudinal expansion, LCMS
correlation functions are nothing more than approximations to fixed
frame correlation functions in rapidity coordinates \cite{chap2}.

Using the same source parameters as in Fig.~1, the effect of the
cross term can be seen in Fig.~2 where we plot the correlator as a
function of ${\rm y}$ for $q_o=30$ MeV. The accuracy of the analytic
approximation (dashed) is seen by comparing it with the exact
numerical result (solid).

We have shown that an ``out-longitudinal'' (or ``out-rapidity'') cross
term arises naturally both in a general gaussian derivation of the
correlation function and in two specific gaussian models.  Although
transformed, in general the cross term persists when one switches from
calculating momentum differences in a fixed frame to calculating them
in the LCMS. Consequently, there is no reason why such a term should
be excluded a priori from gaussian fits to experimental correlation
data. Not only will the new parameter reveal more information about
the source, its inclusion will undoubtedly increase the accuracy of
the other fitted radii.

\acknowledgements

We would like to thank U. Mayer, T. Cs\"org\"o, M. Gyulassy and
Yu. Sinyukov for enlightening discussions. Financially, this work was
supported by BMBF and DFG.


\begin{references}
\bibitem{time}
  This effect arises at a higher order in $(\delta\eta)_*$ than
  was studied in \cite{csorgo}.
\end{references}
\end{document}